\documentclass{article}
\begin{document}
{ \centerline {\bf{Inflation Model(with doublet scalar field) Consistent with ${\Lambda}$-CDM
and WMAP Cosmological Observations}}}

\vskip 0.1cm

{ \centerline {\bf{Amruth B R }}}
{ \centerline {Department of Physics}}
{ \centerline {BITS Pilani, India }}

\vskip 0.1cm

{\centerline {\bf{Ajay Patwardhan}}}
{\centerline { Department of Physics}}
{\centerline {St. Xaviers College, Mumbai,India}}

\vskip 0.1cm

{\centerline {October 2006}}



\begin{abstract}

Cosmological inflation models with modifications to include recent
cosmological observations has been an active area of research after
WMAP 3 results, which have given us information about the
composition of dark matter, normal matter and dark energy and the
anisotropy at the 300000 years horizon with high precision. We work
on inflation models of Guth and Linde and modify them by introducing
a doublet scalar field to give normal matter particles and their
supersymmetric partners which result in normal and dark matter of
our universe. We include the cosmological constant term, as the
vacuum expectation value of the stress energy tensor, as the dark
energy. We calibrate the parameters of our model using recent
observations of density fluctuations. We develop a model which
consistently fits with the recent observations.

\end{abstract}

\section{Introduction}

With new observations about the universe about the dark matter
content and its nature, dark energy and density fluctuations at the
300000 year observable horizon, it has become essential to revisit
the standard cosmological models and incorporate these observations.

In the pre-inflationary epoch, there have been diverse developments
in Loop Quantum Gravity, Strings \& Brane world models, Quantum
Gravity et al. Same is the situation in the post-inflationary epoch
where we have new theories being added in areas of matter-antimatter
models, Baryogenesis, Quark-Gluon models et al. We focus on the
inflationary era and take a fresh look at the standard models, in
particular those of Guth [1] and Linde [2] in the light of these new
findings and the recent cosmological observations from WMAP.

The WMAP results [3] strongly supports the non-baryonic nature of
dark matter, hypothesis about the cosmological constant being the
best candidate for dark energy and the fixing of density
fluctuations at ${10^{-5}}$. These results also suggest that the
${\frac{1}{2}m^2 \phi^2}$ model is a better candidate for inflation
than the ${\frac{1}{4}\lambda \phi^4}$ model [3].

Rapid progress has been seen in incorporating these elements of
observation into theoretical models. Diverse attempts are being made
in this regard as can be seen from the following references of
[6][7][8][9][10].

We start with the standard models and modify them on the basis of
the following three assumptions,
\begin{itemize}
    \item A doublet scalar field drives inflation producing normal
    matter particles and their supersymmetric partners which result
    in dark matter after they acquire mass from the Higgs and
    Higgsino Fields.
    \item Cosmological constant as the vacuum expectation value of
    the stress-energy tensor is the source of dark energy.
    \item The observed density fluctuations are due to quantum
    fluctuations in the primal scalar fields.
\end{itemize}

We use the observational values of parameters about dark matter,
dark energy and the density fluctuations to fix the parameters in
our model. Since we deal with an epoch which is not within the
directly observable horizon, we expect the results of our model to
match with the initial requirements of the post-inflationary epoch
which predicts the observed results at the 300000 year horizon.

\section{Inflationary Cosmology} In the standard model of the
universe with Big-Bang, we start with a singularity, i.e as time ${t
\rightarrow 0}$, the temperature ${T \rightarrow \infty}$. However
we do not yet have a physical framework to deal with such extreme
temperatures and also at such large energy scales and small spacial
scales, quantum gravitational effects become prominent and our
classical theories fail. Therefore we have to resort to new models
to understand the evolution of the universe beyond the planck energy
scale (Planck mass) of ${M_P = \frac{hc}{\sqrt{G}}}$. In units where
${h}$ and ${c}$ are 1, ${M_P = \frac{1}{\sqrt{G}} \simeq 1.22 \times
10^{19} GeV}$.

\subsection{Problems with Classical Big-Bang Model} Since we cannot
deal with extreme temperatures, we start with a temperature lower
than the temperature corresponding to Planck energy scale, ${T_o
\simeq 10^{17} GeV}$ in our standard model and assume that the
initial universe by this time to be homogeneous, isotropic and at
thermal equilibrium at temperature ${T_o}$. Thus, under these
assumptions we can use the Robertson - Walker model to understand
the universe,

\begin{equation}
d\tau^2 = dt^2 - a^2(t)[\frac{dr^2}{1-kr^2} + r^2(d\theta^2 + \sin^2
\theta d\phi^2)]
\end{equation}

where ${a(t)}$ is the scale factor and the three normalized values
of k, ${-1, 0, +1 }$ correspond to open, critical and closed
universe models respectively. The equations of motion for
Friedmann-Robertson-Walker universe can be written as,

\begin{equation}
\ddot{a} = -\frac{4\pi G}{3}(\rho + 3p)a \label{frw-1}
\end{equation}

\begin{equation}
H^2 + \frac{k}{a^2} = \frac{8\pi}{3}G\rho \label{frw2}
\end{equation}

where ${\rho}$ and ${p}$ are the energy density and pressure terms
and ${H}$ is the Hubble parameter,

$ H= \frac{\dot{a}}{a}$

Now, we use the assumption that the universe is filled with massless
particles [1] which are at thermal equilibrium at temperature, say
${T}$, to get the equations for the state of matter:

Energy Density,
\begin{equation}
\rho = 3p = \frac{\pi^2}{30}[N_b(T) + \frac{7}{8}N_f(T)]T^4
\label{thermoenergydensity}
\end{equation}

Entropy,
\begin{equation}
s=\frac{2\pi^2}{45}[N_b(T) + \frac{7}{8} N_f(T)]T^3
\label{thermoentropy}
\end{equation}

Particle number density,
\begin{equation}
n=\frac{1.20206}{\pi^2}[N_b(T) + \frac{3}{4}N_f(T)]T^3
\label{thermonumberdensity}
\end{equation}

where ${N_b(T)}$ and ${N_f(T)}$ are the number of bosonic and
fermionic particles in the universe at that temperature. From now,
on we represent ${[N_b(T) + \frac{7}{8} N_f(T)]}$ by ${N_1}$ and
${[N_b(T) + \frac{3}{4}N_f(T)]}$ by ${N_2}$.

Now, we substitute ${\rho}$ from Eq.~(\ref{thermoenergydensity})
into Eq.~(\ref{frw2}) and express ${H}$ as ${(\frac{\dot{T}}{T})}$ to
get

\begin{equation}
(\frac{\dot
{T}}{T})^2 + \frac{k}{a^2 T^2} T^2 = \frac{8\pi}{3} G
\frac{\pi^2}{30} N_1 T^4
\end{equation}
i.e,
\begin{equation}
(\frac{\dot{T}}{T})^2 + \epsilon(T) T^2 = \frac{4}{45}\pi^3 G N_1 T^4
\end{equation}

where ${\epsilon(T) = \frac{k}{a^2 T^2}}$.

This can be expressed as,

\begin{equation}
\epsilon(T)=k[\frac{1}{a^3 T^3}]^{\frac{2}{3}}
\end{equation}
i.e,
\begin{equation}
\epsilon(T) = k[\frac{\frac{2\pi^2}{45} N_1}{a^3 \frac{2\pi^2}{45}
N_1 T^3}]^{\frac{2}{3}}
\end{equation}
Using Eq.~(\ref{thermoentropy}), this becomes
\begin{equation}
\epsilon(T) = k[\frac{2\pi^2 N_1}{45 a^3 s}]^{\frac{2}{3}}.
\end{equation}

But ${a^3 s}$ is the total entropy in a volume of radius ${a}$, the
scale factor,

\begin{equation}
S=a^3 s .
\end{equation}
Therefore
\begin{equation}
\epsilon(T) = k[\frac{2\pi^2 N_1}{45 S}]^{\frac{2}{3}}.
\end{equation}

\subsubsection{Adiabatic Universe Approximation}

This approximation uses the conservation of entropy in the early
universe and gives the expression for the energy density of the
universe as,

\begin{equation}
|\frac{\rho - \rho_{cr}}{\rho}| = \frac{45}{4\pi^3}\frac{M_P ^2}{N_1
T^2} |\epsilon|.
\end{equation}

where ${\rho_{cr}}$ is the critical energy density of the universe.

To explain the evolution of the universe to its current state, we
require the total entropy, ${S>10^{86}}$. This condition gives

\begin{equation}
|\epsilon(T)| < k[\frac{2\pi^2 N_1}{45 \times
10^{86}}]^{\frac{2}{3}} \label{epsilonadia}
\end{equation}
This constrains the energy density equation as
\begin{equation}
|\frac{\rho - \rho_{cr}}{\rho}| < k^{"} 10^{-58} N_1^{-\frac{1}{3}}
\frac{M_P ^2}{T^2} \label{flatnessprob}
\end{equation}

where ${k^{"}}$ is another constant. Such a small value in the RHS
implies that the energy density of the universe was incredibly close
to the critical energy density of the universe, at the beginning of
Big-Bang, in order to explain the current evolution of the universe.
This remarkable fine-tuning of energy density necessary, is known as
the \textbf{flatness problem}.

Now, let us consider how causally well connected the early universe
was. We represent the radius of the physical horizon, i.e of the
region which could be causally connected, by ${l(t)}$. The radius of
the universe required to evolve into its current state is
represented by ${L(t)}$. The ratio of the corresponding volumes
comes out to be,

\begin{equation}
\frac{l^3}{L^3} \simeq 4 \times 10^{-89} N_1 ^{-\frac{1}{2}}
(\frac{M_P}{T})^3. \label{horizonprob}
\end{equation}

where from entropy conservation, we require ${L^3 s = L_P ^3 s_P}$,
 ${L_P}$ and ${s_P}$ being present radius and entropy.

Now, the RHS of Eq.~(\ref{horizonprob}) being so small, implies that
very small portion of the early universe was causally connected.
Thus the energy density could not have undergone redistribution and
hence would be non-uniform. This would further imply that the cosmic
microwave background radiation should have been highly non-uniform
which contradicts the observations. This is the \textbf{horizon
problem} of classical Big-Bang model.

\subsubsection{Non-Adiabatic Early Universe} Now, let us remove the
adiabatic assumption and allow for entropy generation in early
universe given by

\begin{equation}
S_P = Z^3 S_o
\end{equation}

where ${S_o = a^3 s}$ is the initial entropy and ${Z}$ is the scale
factor for entropy generation. With this, Eq.~(\ref{epsilonadia})
becomes

\begin{equation}
|\epsilon(T)| < k[\frac{2\pi^2 N_1}{45 \times
\frac{10^{86}}{Z^3}}]^{\frac{2}{3}}
\end{equation}
This modifies Eq.~(\ref{flatnessprob}) to
\begin{equation}
|\frac{\rho - \rho_{cr}}{\rho}| < k^{"} Z^2 10^{-58}
N_1^{-\frac{1}{3}} \frac{M_P ^2}{T^2}
\end{equation}

For ${|\frac{\rho - \rho_{cr}}{\rho}|}$ to be within the current
observational range, ${Z}$ needs to be of the order of ${10^{27}}$.
Now similarly we have

\begin{equation}
L^3 Z^3 s = L_P ^3 s_P
\end{equation}

because of which Eq.~(\ref{horizonprob}) will now be

\begin{equation}
\frac{l^3}{L^3} \simeq Z^3 \times 4 \times 10^{-89} N_1
^{-\frac{1}{2}} (\frac{M_P}{T})^3.
\end{equation}

Again for the RHS to be approximately equal to one which would mean
a causally connected early universe, we require ${Z}$ to be of the
order ${10^{27}}$.

\section{Classical Inflation} In our non-adiabatic early universe
approximation, we saw that a large value for the scale factor of
entropy production solves the flatness and horizon problem.
Classical inflation [1] provides such a model where such large
entropy generation is possible.

\subsubsection{Inflationary Universe} We start with the assumption
that the equations for state of matter in early universe exhibit a
first order phase transition at some critical temperature, ${T_c}$.
Then just like any phase diagram of thermodynamics would suggest, we
can expect bubbles of low temperature phase, say phase-2 to nucleate
and grow in a universe of initial state, phase-1. Now, if the
nucleation rate for this transition is very low, the universe will
not undergo a transition at constant temperature but instead cools
further due to expansion. It will then be supercooled phase-2
bubbles in high-temperature phase-1 universe. If this supercooled
temperature, ${T_s}$ is many orders below ${T_c}$, then when
phase-transition finally takes place at ${T_s}$ the latent heat
released will be huge relative to ${T_s}$ since latent heat is a
characteristic of the energy scales at ${T_c}$, the critical
temperature of transition. This reheats the universe to some
temperature ${T_r}$ comparable to ${T_c}$. This causes an entropy
increase by a factor of ${(\frac{T_r}{T_s})^3}$ while the scale
factor of the universe, ${a}$ remains unchanged. Thus in this model,
we have
\begin{equation}
Z=\frac{T_r}{T_s}.
\end{equation}
For this model to solve the flatness problem and the horizon
problem, ${Z}$ needs to be of the order of ${10^{27}}$ and hence the
universe should supercool by 27 exponential orders or more magnitude
of temperature below ${T_c}$.

\subsection{Evolution of Universe in Classical Inflation} Classical
inflation explains the supercooling of the early universe by
proposing a model in which the early universe was in a state of
false vacuum characterized by a scalar field in a local minimum of
its potential energy function. In such a model, as the temperature
drops and ${T \rightarrow 0}$, the universe cools not towards true
vacuum but towards the false vacuum state of energy density
${\rho_o}$, which is greater than the true vacuum energy density.

Therefore Eq.~(\ref{thermoenergydensity}) for energy density becomes

\begin{equation}
\rho(T) = \frac{\pi^2}{30} N_1 T^4 + \rho_o.
\end{equation}
Therefore Eq.~(\ref{frw2}) now takes the form
\begin{equation}
(\frac{\dot{T}}{T})^2 + \epsilon(T)T^2 = \frac{4\pi^3}{45}G N_1 T^4 +
\frac{8\pi}{3}G \rho_o. \label{frw-2cic}
\end{equation}

Based on the value of ${\epsilon}$ this equation will have two types
of solution. If ${\epsilon > \epsilon_o}$, where

\begin{equation}
\epsilon_o = \frac{8\pi^2}{45}\sqrt{30} G\sqrt{N_1 \rho_o}
\end{equation}
Then the universe will have an expanding phase which halts when
temperature reaches ${T_{min}}$ given by
\begin{equation}
T_{min}^4 = \frac{30}{\pi^2}\rho_o [\frac{\epsilon}{\epsilon_o} +
\{(\frac{\epsilon}{\epsilon_o})^2 - 1\}^{\frac{1}{2}}]^2
\end{equation}
After this, the universe starts to contract.

For the case when ${\epsilon < \epsilon_o}$, we can consider it to
mainly be ${\epsilon < 0}$ since ${\epsilon_o \simeq 0}$. In this
case we get

\begin{equation}
T(t) = Const. e^{-\chi t},
\end{equation}
where
\begin{equation}
\chi^2 = \frac{8\pi}{3}G\rho_o.
\end{equation}

But since ${aT=constant}$, we get
\begin{equation}
a(t) = Const. e^{+\chi t}.
\end{equation}

Since this will be the universe which will reach observable sizes,
we are interested in the second solution which has an inflationary
phase of exponential growth. Thus we get a universe which is
exponentially expanding in a false vacuum state of energy density
${\rho_o}$. For considerable flattening of the universe, we need an
exponential expansion of at least order 60.

\subsubsection{Shortcomings of Classical Inflation}
Classical Inflation is based on several assumptions which do not
come naturally in the model. The nucleation rate needs to be very
slow compared to the expansion rate of the universe in order for
${Z}$ to be of the order ${10^{27}}$. Also, the randomness of bubble
formation in analogy with thermodynamic situations should have led
to inhomogeneities in the universe.

Also, the state of false vacuum would be stable in classical theory
since there would be no energy available to allow the scalar field
to cross the potential barrier that separates it from the lower
energy states. However this state would decay through quantum
mechanical tunneling and thereby ending inflation. Since this decay
of false vacuum due to quantum tunneling is random, it would lead to
large inhomogeneities in the universe.

Further, when classical inflation was proposed, the dark-matter and
dark-energy problems were not linked to cosmological model. However
it has now become necessary for inflation to answer the puzzles
about the production and nature of dark-matter as well as dark
energy.

\section{Chaotic Inflation} Chaotic Inflation was proposed to solve
the problems faced by classical inflation [2]. It assumes that the
initial state of the universe was in a state of quantum chaos. If
this chaotic state can produce a large enough scalar field ${\phi}$
then inflation can occur and bring about the same effects which
classical inflation does, as the scalar field relaxes from its
initial state.

\subsection{Evolution of Universe in Chaotic Inflation - ${\lambda
\phi^4}$ model}

Here we us consider a scalar field potential [4] given by
\begin{equation}
V(\phi)=\frac{1}{4}\lambda \phi^4.
\end{equation}
The energy density will be proportional to ${(\frac{\partial
\phi}{\partial x^{\mu}})^2}$, with ${x^{\mu}}$ representing the four
co-ordinates.

A classical description of the universe is possible only after the
energy density becomes smaller than ${M_P^4}$, i.e after planck time
of ${\frac{1}{M_P}}$, where ${M_P=\frac{1}{\sqrt{G}}}$. This
constraint implies that the potential at ${t=\frac{1}{M_P}}$ should
satisfy

\begin{equation}
V(\phi) \leq M_P^4
\end{equation}
i.e
\begin{equation}
-(\frac{4}{\lambda})^{\frac{1}{4}} M_P \leq \phi \leq
+(\frac{4}{\lambda})^{\frac{1}{4}} M_P
\end{equation}

Before this time, the universe is assumed to be in a chaotic quantum
state.

Let us now consider a locally homogeneous region in such early
universe. This part is treated as expanding De-Sitter space with
scale factor ${a(t)=a_o e^{Ht}}$ with

\begin{equation}
H=\sqrt{\frac{8\pi GV}{3}}
\end{equation}
From the equations of motion in De-Sitter space, we have
\begin{equation}
\frac{\partial^2 \phi}{\partial t^2} + 3H\frac{\partial
\phi}{\partial t} =-\frac{\partial V}{\partial \phi} = -\lambda
\phi^3.
\end{equation}
Assuming that ${\frac{M_P^2}{6\pi}}$ i.e ${\frac{1}{6\pi G} \ll
\phi^2}$, we write the RHS as,
\begin{equation}
-\lambda \phi^3 = \lambda \phi(-\phi^2) \approx \lambda \phi
(\frac{M_P^2}{6\pi} - \phi^2)
\end{equation}
\begin{equation}
-\lambda \phi^3 \approx \frac{\lambda \phi M_P^2}{6\pi} - \lambda
\phi^3
\end{equation}

And the LHS can be written as
\begin{equation}
\frac{\partial^2 \phi}{\partial t^2} + 3H\frac{\partial
\phi}{\partial t} = \frac{\partial^2 \phi}{\partial t^2} +
3\sqrt{\frac{2\pi \lambda}{3}} \frac{\phi^2}{M_P} \frac{\partial
\phi}{\partial t}
\end{equation}
i.e
\begin{equation}
\frac{\partial^2 \phi}{\partial t^2} + 3H\frac{\partial
\phi}{\partial t} = \frac{\partial^2 \phi}{\partial t^2} +
\sqrt{\frac{6\pi \lambda}{M_P^2}} \phi^2 \frac{\partial
\phi}{\partial t}.
\end{equation}

Equating the LHS and the RHS, we get
\begin{equation}
\frac{\partial^2 \phi}{\partial t^2} + \sqrt{\frac{6\pi
\lambda}{M_P^2}} \phi^2 \frac{\partial \phi}{\partial
t}=\frac{\lambda \phi M_P^2}{6\pi} - \lambda \phi^3
\end{equation}
Equating the second term on both sides gives both derivatives of
${\phi}$ in terms of ${\phi}$, indicating an exponential behaviour
of ${\phi}$
\begin{equation}
\sqrt{\frac{6\pi \lambda}{M_P^2}} \phi^2 \frac{\partial
\phi}{\partial t}=-\lambda \phi^3
\end{equation}
which gives the solution as
\begin{equation}
\phi = \phi_o e^{-\frac{\sqrt{\lambda}M_P}{\sqrt{6\pi}} t}
\end{equation}
Putting this in ${\frac{\partial^2 \phi}{\partial t^2}}$, we can
verify that
\begin{equation}
\frac{\partial^2 \phi}{\partial t^2} = \phi_o
(\frac{\sqrt{\lambda}M_P}{\sqrt{6\pi}})^2
e^{-\frac{\sqrt{\lambda}M_P}{\sqrt{6\pi}} t}
\end{equation}
i.e,
\begin{equation}
\frac{\partial^2 \phi}{\partial t^2} = \frac{\lambda \phi
M_P^2}{6\pi}.
\end{equation}

Thus we take ${\phi=\phi_o e^{-\frac{\sqrt{\lambda}M_P}{\sqrt{6\pi}}
t}}$ as a solution. Here we used the assumption of ${\phi^2 \gg
\frac{1}{6\pi G}}$ which is valid only when ${\lambda \ll 1}$ and
${V(\phi) \leq M_P^4}$.

From our solution we see that ${\phi}$ has a time constant of ${t_c
= \frac{\sqrt{6\pi}}{M_P \sqrt{\lambda}}}$. Therefore in one time
constant the universe expands by

\begin{equation}
a(t_c) = a_o e^{Ht_c}
\end{equation}

\begin{equation}
a(t_c)=a_o e^{\phi_o^2 e^{-2}\sqrt{\frac{2\pi \lambda}{3 M_P^2}}
\sqrt{\frac{6\pi}{M_P^2 \lambda}}}
\end{equation}

\begin{equation}
a(t_c) \simeq a_o e^{\frac{2\pi \phi_o^2}{M_P ^2}}.
\label{lambdaphiexp}
\end{equation}

Now, for inflation to flatten the universe considerably in this
period we require an expansion of at least 60 e-foldings, i.e
${a(t_c)\geq a_o e^{60}}$. Therefore

\begin{equation}
\frac{2\pi \phi_o^2}{M_P^2} \geq 60
\end{equation}
i.e,
\begin{equation}
\phi_o \geq 3.09 M_P.
\end{equation}

Now considering the two constraints which we derived in our model,
\begin{equation}
\phi \leq (\frac{4}{\lambda})^{\frac{1}{4}}M_P
\end{equation}
and
\begin{equation}
\phi_o \geq 3.09 M_P
\end{equation}
 we see that ${\lambda}$ should be very small for both the
 constraints to be satisfied. Taking ${\phi \approx \phi_o}$, we
 approximate ${\lambda}$ as
 \begin{equation}
 3.1 M_P \leq (\frac{4}{\lambda})^{\frac{1}{4}} M_P
 \end{equation}
 \begin{equation}
 \Rightarrow \lambda < 0.04
 \end{equation}

 The rapid exponential rate suggests that once such a patch of
 homogeneous and isotropic field is formed, it expands rapidly to
 dominate the physical volume of the universe. This naturally
 accounts for the homogeneity of the observed universe.

 The energy density at the end of this initial inflationary phase
 will be

 \begin{equation}
 \rho_o \approx \frac{1}{4}\lambda \phi_o^4 > 22.82 \lambda M_P^4
 \end{equation}
 Therefore for ${\lambda \approx 0.04}$ we have
 \begin{equation}
 \rho_o \approx 0.91 M_P^4 < M_P^4
 \end{equation}
 thereby satisfying our constraint on the energy density.

\subsection{Evolution of Universe in Chaotic Inflation - ${m^2
\phi^2}$ model} Since the recent WMAP results [3] have favoured the
${\frac{1}{2}m^2 \phi^2}$ model, we now consider a potential of the
form

\begin{equation}
V(\phi)=\frac{1}{2}m^2 \phi^2
\end{equation}
as the starting potential in the early universe [4]. We use
${V(\phi) \leq M_P^4}$ to get
\begin{equation}
-(\frac{2}{m^2})^{\frac{1}{2}}M_P^2 \leq \phi \leq +
(\frac{2}{m^2})^{\frac{1}{2}}M_P^2. \label{mphiconstraint1}
\end{equation}

Proceeding in the same way as we did for the ${\lambda \phi^4}$
model, we get the equation of motion as
\begin{equation}
\frac{\partial^2 \phi}{\partial t^2} + 3H\frac{\partial
\phi}{\partial t} =-\frac{\partial V}{\partial \phi} = - m^2 \phi.
\end{equation}
Substituting for ${H}$ as
\begin{equation}
H = \sqrt{\frac{8\pi G V}{3}} = \sqrt{\frac{8\pi G m^2 \phi^2}{6}}
\end{equation}
we get
\begin{equation}
\frac{\partial^2 \phi}{\partial t^2} + \sqrt{12 \pi G} m \phi
 \frac{\partial \phi}{\partial t}= - m^2 \phi
\end{equation}
which can be solved to get ${\phi}$ as
\begin{equation}
\phi = \phi_o - \frac{m}{\sqrt{12 \pi G}} t.
\end{equation}

Now, since we have ${a(t) = a_o e^{Ht}}$, we calculate how much the
universe has expanded by the time the scalar field relaxes to 0 as
\begin{equation}
a(t) = a_o e^{N}
\end{equation}
where the exponent ${N}$ is given by
\begin{equation}
N = \int_{\phi=\phi_o}^{\phi = 0}{H(t) dt = 2 \pi G \phi_o^2.}
\end{equation}

Since we require the universe to have undergone an expansion of at
least 60 e-foldings, we require ${N \geq 60}$. Therefore
\begin{equation}
\phi_o \geq \sqrt{\frac{60}{2 \pi G}}. \label{mphiconstraint2}
\end{equation}
We compare Eq.~(\ref{mphiconstraint1}) and
Eq.~(\ref{mphiconstraint2}) by setting ${\phi \approx \phi_o}$ to
get
\begin{equation}
\sqrt{\frac{60}{2 \pi G}} \leq (\frac{2}{m^2})^{\frac{1}{2}}M_P^2
\end{equation}
giving the limit on ${m}$ as
\begin{equation}
m \leq \sqrt{\frac{4\pi}{60 G}}.
\end{equation}

Thus we again have a theory which allows for an inflationary phase
in the evolution of the universe. However we see here that the
condition on ${m}$ is not as stringent as it was on ${\lambda}$ in
the ${\lambda \phi^4}$ model.

\section{Inflationary Model Consistent with ${\Lambda}$-CDM \& WMAP}

After WMAP results [3 - Section 6] have been published, revisions of
inflationary models are being actively published. Some of the
attempts inspect the inclusion of dark matter and dark energy based
on WMAP results [14][6]. Our model also contrasts with [6][8] in
which the goal of including dark matter, dark energy and
inhomogeneity is addressed in a different way. We develop a model
consistent with the recent cosmological observations and based on
the hypothesis from ${\Lambda}$-CDM model.

\subsection{Overview of modifications to standard models} We work
with a pair ${\phi_o, \bar{\phi_o}}$ of primal scalar fields which
transform to ${\phi_1, \phi_2}$ by

\begin{equation}
\left(%
\begin{array}{c}
  \phi_1 \\
  \phi_2 \\
\end{array}%
\right) = \left(%
\begin{array}{cc}
  \cos \theta_{SS} & \sin \theta_{SS} \\
  -\sin \theta_{SS} & \cos \theta_{SS} \\
\end{array}%
\right) \left(%
\begin{array}{c}
  \phi_o \\
  \bar{\phi_o} \\
\end{array}%
\right)
\end{equation}

such that the action is defined by the potential,

\begin{equation}
V = V_o - \frac{1}{2}\mu^2 \phi^{\dag} \phi + \frac{1}{4}\lambda
(\phi^{\dag}\phi)^2
\end{equation}
where
\begin{equation}
\phi^{\dag} \phi = \left(%
\begin{array}{cc}
  \phi_o^- & \phi_o \\
\end{array}%
\right) \left(%
\begin{array}{c}
  \phi_o \\
  \bar{\phi_o} \\
\end{array}%
\right) = \left(%
\begin{array}{cc}
  \phi_1^* & \phi_2^* \\
\end{array}%
\right) \left(%
\begin{array}{c}
  \phi_1 \\
  \phi_2 \\
\end{array}%
\right)
\end{equation}
Since this is dependent only on even power terms, invariance of even
power terms after transformation will ensure that the universe
evolves in an identical way.

In this formulation, we have exact SUSY for
${\theta_{SS}=\frac{\pi}{4}}$. As ${\theta_{SS} \rightarrow 0}$ or
${\frac{\pi}{2}}$, we have broken SUSY with one or the other field
preferred. The correct ${\theta_{SS}}$ resulting in partially broken
SUSY will be obtained from matching the derived results with
observations.

In this model, we intend to show that normal matter and radiation
are generated by ${\phi_2}$ and the dark matter by ${\phi_1}$. [This
step is guided by numerous examples in physics where we have
doubling and mixing of particles and fields such as $(\kappa_o,
\bar\kappa_o)$,$(\nu_e, \nu_{\mu})$, W-S model, particle-antiparticle
et al due to the breaking of some underlying symmetry.] Here, the
${\phi_1 - \phi_2}$ doublet due to broken SUSY will have different
masses, numbers and lifetimes of their particles. So ${\phi_2}$
gives rise to normal matter particles and ${\phi_1}$ to their
supersymmetric partners. This model has a ${\theta_{SS}}$ parameter
in addition to ${\mu}$ and ${\lambda}$ with which the observed ratio
of normal matter to dark matter, ${1:5}$ and the production rate of
supersymmetric particles is handled.

The dark energy problem is tackled by identifying it with the
cosmological constant ${\Lambda}$ of the universe which is taken as
the vacuum expectation values of the stress-energy tensor. Further,
${\langle T_{\mu \nu} \rangle_{vac} \equiv \Lambda g_{\mu \nu}}$ for
the ${p=-\rho}$ equation of state for dark energy as they are the
only locally lorentz covariant form consistent with the recently
found accelerating phase of the universe [12][13].

The classical cosmological models use a constant field ${\phi}$ for
homogeneous, isotropic models of the universe. However, the
fluctuation in the scalar field, ${\phi \rightarrow \phi + \delta
\phi}$ may give rise to anisotropy and inhomogeneity as observed in
COBE and WMAP data viz ${\frac{dT}{T}\approx \frac{dp}{p} \approx
\frac{d\rho}{\rho} \approx 10^{-5}}$. To do this classically we need
a stochastic inflation model. But as ${E}$ evolves from ${10^{19}}$
GeV after ${10^{-43}}$s from Big-Bang to ${10^3}$ GeV corresponding
to the QGP baryogenesis at ${10^2}$s after Big-Bang, we have ${E
\approx pc}$ for most particles and hence ${\lambda_{DB} \approx
\frac{h}{p} \approx \frac{hc}{E}}$. So ${\lambda_{DB}}$ is
comparable to the size of the universe during inflation. Hence
quantum effects should dominate and the fluctuations should be
calculated quantum mechanically viz for any observable A,

\begin{equation}
\triangle \hat{A} = [ \langle A^2 \rangle - \langle A
\rangle^2]^{\frac{1}{2}}
\end{equation}
For our purpose, we will be handling the average values for the
number operator and the hamiltonian. Hence the observed fluctuations
at 3,00,000 year observable horizon must arise from the ratio
${\frac{\triangle N}{N}}$ and ${\frac{\triangle H}{H}}$ found.
[These fluctuations are correlated with the large scale structure
formation of galaxies, voids and cluster filaments as seen in early
universe. These fluctuations are also expected to be the remnant
effects of gravity wave spectrum that arose from Big-Bang and is to
be confirmed by the Planck satellite].

Some of the parameters in our model, namely ${\mu, \lambda}$ and
${\theta_{SS}}$ can also be fixed using this observed and calculated
fluctuations. Our model also contrasts with [6][8] in which the goal
of including dark matter, dark energy and inhomogeneity is addressed
in a different way.

\subsection{The Scalar Field Doublet Model for Inflation}

We start with a complex pair of primal scalar fields ${\phi_o,
\bar{\phi_o}}$ which could be the remnant of the pre-inflationary epoch. We then define a doublet $\left(%
\begin{array}{c}
  \phi_1 \\
  \phi_2 \\
\end{array}%
\right)$ as the driving fields for inflation and generated by a
transform that mixes ${\phi_o,\phi_o^{-}}$ as inflation proceeds,

\begin{equation}
\left(%
\begin{array}{c}
  \phi_1 \\
  \phi_2 \\
\end{array}%
\right) = \left(%
\begin{array}{cc}
  \cos \theta_{SS} & \sin \theta_{SS} \\
  -\sin \theta_{SS} & \cos \theta_{SS} \\
\end{array}%
\right) \left(%
\begin{array}{c}
  \phi_o \\
  \bar{\phi_o} \\
\end{array}%
\right).
\end{equation}

Now calculating the energy terms for these new fields, we find that

\begin{equation}
\phi_1^{\dag} \phi_1 = \left(%
\begin{array}{cc}
  \cos \theta_{SS} \phi_o^- & \sin \theta_{SS} \phi_o \\
\end{array}%
\right)\left(%
\begin{array}{c}
  \cos \theta_{SS} \phi_o \\
  \sin \theta_{SS} \bar{\phi_o} \\
\end{array}%
\right)
\end{equation}
which reduces to
\begin{equation}
\phi_1^{\dag} \phi_1 = (\cos^2 \theta_{SS} + \sin^2
\theta_{SS})\phi_o \bar{\phi_o} = \phi_o \bar{\phi_o}.
\end{equation}
Similarly we have for ${\phi_2}$
\begin{equation}
\phi_2^{\dag} \phi_2 = \phi_o \bar{\phi_o}.
\end{equation}

Thus we see that this transform ensures that the energy terms remain
the same as the original complex field ensuring that inflation
proceeds in the same way as it would in a model without such a
symmetry breaking, but now with a doublet scalar field $\left(%
\begin{array}{c}
  \phi_1 \\
  \phi_2 \\
\end{array}%
\right)$.

Starting with these two fields, we go on to define the potential in
the early universe as

[Case:1]
\begin{equation}
V_1 = \frac{1}{4} \lambda_1 \phi_1^4
\end{equation}
\begin{equation}
V_2 = \frac{1}{4} \lambda_2 \phi_2^4.
\end{equation}
These potentials when used in the standard ${\lambda \phi^4}$
inflationary model give the solution as
\begin{equation}
\phi_1 = \phi_{1_o} e^{-\frac{\sqrt{\lambda_1}M_P}{\sqrt{6 \pi}} t}
\end{equation}
and
\begin{equation}
\phi_2 = \phi_{2_o} e^{-\frac{\sqrt{\lambda_2}M_P}{\sqrt{6 \pi}} t}
\end{equation}
respectively. Here we observe that the rate at which fields decay is
proportional to ${\sqrt{\lambda}}$, i.e
\begin{equation}
\frac{\partial \phi_i}{\partial t} \propto \sqrt{\lambda_i} \ \ \ \
\ i = 1,2.
\end{equation}

[Case:2]
\begin{equation}
V_1 = \frac{1}{2} m_1^2 \phi_1^2
\end{equation}
\begin{equation}
V_2 = \frac{1}{2} m_2^2 \phi_2^2.
\end{equation}
Again using these potentials in the standard ${m^2 \phi^2}$ model,
we obtain the solutions
\begin{equation}
\phi_1 = \phi_{1_o} - \frac{m_1}{\sqrt{12 \pi G}} t
\end{equation}
and
\begin{equation}
\phi_2 = \phi_{2_o} - \frac{m_2}{\sqrt{12 \pi G}} t
\end{equation}
respectively. Here we have the rate at which fields decay
proportional to ${m}$, i.e
\begin{equation}
\frac{\partial \phi_i}{\partial t} \propto m_i \ \ \ \ \ i = 1,2.
\end{equation}

This decay of the field results in the creation of massless
particles that are produced during the inflationary phase.
Accordingly, in our model we have ${\phi_2}$ generating the normal
matter particles and ${\phi_1}$ giving rise to their supersymmetric
partners. Based on recent observations and the ${\Lambda - CDM}$
model, we go by the hypothesis that the dark matter content in our
universe is predominantly due to massive supersymmetric partners, in
particular neutralino which is the lightest and most stable of the
supersymmetric particles. We can therefore relate the decay rate of
${\phi_1}$ and ${\phi_2}$ to the creation rate, ${\Gamma}$ of dark
matter (DM) and normal matter (NM) respectively,
\begin{equation}
\Gamma_{DM, \  NM} \  \propto \ \frac{\partial }{\partial t}
\phi_{1, \ 2}.
\end{equation}

Now, in order to fix the values of the coupling coefficients in our
model, we calculate the predicted density fluctuation and use the
observed value of ${\approx 10^{-5}}$ from the recent WMAP results.
These fluctuations can be evaluated using either stochastic models,
second quantization technique or by using the path integral ensemble
average.

We use the second quantization approximation for the density
fluctuation, ${\delta_H}$ to obtain a limit on these coefficients
[31],

\begin{equation}
\delta_H \approx \frac{H^2}{2\pi \dot{\phi}}
\end{equation}

Since we require the density fluctuation for normal matter, we
evaluate it for ${\phi_2}$. For the ${\lambda \phi^4}$ model, we
have

\begin{equation}
\dot{\phi_2} = \frac{-\lambda_2 \phi_2^3}{3 H}
\end{equation}
Therefore
\begin{equation}
\delta_H \approx \frac{3H^3}{2 \pi \lambda_2 \phi_2^3}
\end{equation}
which reduces to
\begin{equation}
\delta_H \approx \frac{1}{2\pi}
\sqrt{\frac{\lambda_2}{3}}(\frac{2\pi
\phi_2^2}{M_P^2})^{\frac{3}{2}} \approx 10^{-5}
\end{equation}

With ${\phi_2 \approx \phi_{2_o}}$, we see that the term
${(\frac{2\pi \phi_2^2}{M_P^2})}$ is the exponent in
Eq.~(\ref{lambdaphiexp}) and therefore needs to be ${\geq 60}$. This
gives the limit for ${\lambda_2}$ as
\begin{equation}
\lambda_2 \leq \frac{12\pi^2 \times 10^{-10}}{60^3}
\end{equation}
\begin{equation}
\lambda_2 \leq 54.83 \times 10^{-15}. \label{lambdalimit2}
\end{equation}

Similarly, for the ${m^2 \phi^2}$ model, we get

\begin{equation}
\dot{\phi_2} = \frac{-m_2^2 \phi_2}{3 H}
\end{equation}
which gives the density fluctuation as
\begin{equation}
\delta_H \approx \frac{6m_2}{M_P \pi^2}(\frac{\pi}{3})^{\frac{3}{2}}
2\pi G \phi_2^2
\end{equation}
Again we have ${2\pi G \phi_2^2 \geq 60}$ which constraints ${m_2}$
as
\begin{equation}
m_2 \leq 2.5583 \times 10^{-7} M_P. \label{mlimit2}
\end{equation}

Now, we use our hypothesis that ${\phi_1}$ generates supersymmetric
particles that constitute dark matter and ${\phi_2}$ generates
normal matter particles. These particles produced during inflation
acquire mass because of their interaction with the Higgs Field which
gives mass to normal particles and the Higgsino Field which gives
mass to supersymmetric particles, in the post-inflationary epoch. We
take the standard normal matter particle as a proton which has an
energy of 1 GeV and the standard supersymmetric particle as the
neutralino with energy around 100 GeV. We now estimate the ratio of
mass of dark matter to the mass of normal matter as

\begin{equation}
\frac{M_{DM}}{M_{NM}} \ \propto \ \frac{\Gamma_{DM} \times
M_{neutralino}}{\Gamma_{NM} \times M_{proton}}
\end{equation}

For ${\lambda \phi^4}$ model, this becomes
\begin{equation}
\frac{M_{DM}}{M_{NM}} \ \propto \ \frac{\sqrt{\lambda_1} \times 100
\ GeV}{\sqrt{\lambda_2} \times 1 \ GeV}
\end{equation}

Recent observations of WMAP give us the ${M_{DM} : M_{NM}}$ ratio as
${22:4}$. Therefore

\begin{equation}
\sqrt{\lambda_1} = \frac{22}{400} \sqrt{\lambda_2}
\end{equation}
Using Eq.~(\ref{lambdalimit2}), this limits ${\lambda_1}$ as
\begin{equation}
\lambda_1 \leq 16.586 \times 10^{-17}.
\end{equation}

Similarly for ${m^2 \phi^2}$ model, we get
\begin{equation}
\frac{M_{DM}}{M_{NM}} \ \propto \ \frac{m_1 \times 100 \ GeV}{m_2
\times 1 \ GeV}
\end{equation}
Therefore we get from Eq.~(\ref{mlimit2})
\begin{equation}
m_1 \leq 1.407065 \times 10^{-8} M_P.
\end{equation}

Now that we have fixed the limits for ${\lambda_1, \lambda_2}$ /
${m_1, m_2}$ in our model, we have a model involving a doublet
scalar field, driving inflation and resulting in the production of
normal particles and their supersymmetric partners in the right
ratio to agree with the observed ratio of dark matter to normal
matter. The remnant of these fields can be considered to transmute
into the Higgs and Higgsino fields after the inflationary epoch when
the energy scale of the universe falls below a critical value.

\subsubsection{Source of Density Fluctuations}
As the universe evolves from the scale of ${10^{-35}}$m of initial
size at ${10^{-43}}$s, we see that such small scales inevitably lead
to quantum effects dominating the evolution of the scalar field.
Therefore we expect quantum fluctuations in the scalar field leading
to fluctuations in the hamiltonian which gets expressed as the
fluctuations in the number density of the particles produced during
inflation. When these particles get mass in the post-inflationary
epoch, these fluctuations get manifested as density fluctuations.

Since we are dealing with massless particles, we take only the
potential in evaluating the expectation value of the
hamiltonian,${\langle H \rangle}$.

For the ${\lambda \phi^4}$ model,
\begin{equation}
\langle H \rangle = \int{D[\phi] \frac{1}{4} \lambda \phi^4
e^{-\frac{1}{4}\lambda \phi^4}}
\end{equation}
Making the substitution ${\frac{1}{4} \lambda \phi^4 = y}$, we can
reduce the integral to
\begin{equation}
\langle H \rangle =
\frac{\lambda}{16}(\frac{4}{\lambda})^{\frac{5}{4}} \int{dy
y^{\frac{1}{4}}e^{-y}}
\end{equation}
i.e
\begin{equation}
\langle H \rangle =
\frac{\lambda}{16}(\frac{4}{\lambda})^{\frac{5}{4}}
\Gamma(1+\frac{1}{4}).
\end{equation}
Similarly, we get ${\langle H^2 \rangle}$ by making similar
substitution and simplifying the integral,
\begin{equation}
\langle H^2 \rangle = \int{D[\phi] \frac{1}{16} \lambda^2 \phi^8
e^{-\frac{1}{4}\lambda \phi^4}}
\end{equation}
\begin{equation}
\langle H^2 \rangle =
\frac{\lambda^2}{64}(\frac{4}{\lambda})^{\frac{9}{4}}
\Gamma(1+\frac{5}{4}).
\end{equation}
Now we use ${\langle \triangle H \rangle = [\langle H^2 \rangle -
(\langle H \rangle)^2]^{\frac{1}{2}}}$ to obtain the value of
${\frac{\langle \triangle H \rangle}{\langle H \rangle}}$ as

\begin{equation}
\frac{\langle \triangle H \rangle}{\langle H \rangle} =
\frac{1}{4^{\frac{1}{8}}}\sqrt{5.5163 \lambda^{\frac{1}{4}} -
4^{\frac{1}{4}}}.
\end{equation}
This allows us to limit the value of ${\lambda}$ in the same way as
second quantization scheme.

Similarly for ${m^2 \phi^2}$ model, we evaluate ${\langle H
\rangle}$ as
\begin{equation}
\langle H \rangle = \int{D[\phi] \frac{1}{2} m^2 \phi^2
e^{-\frac{1}{2} m^2 \phi^2}}
\end{equation}
Making the substitution ${\frac{1}{2} m^2 \phi^2 = y}$, we can
reduce the integral to
\begin{equation}
\langle H \rangle = \frac{m^2}{4}(\frac{2}{m^2})^{\frac{3}{2}}
\Gamma(1+\frac{1}{2})
\end{equation}
and similarly ${\langle H^2 \rangle}$ becomes
\begin{equation}
\langle H^2 \rangle =
\frac{m^4}{8}(\frac{2}{m^2})^{\frac{5}{2}}\Gamma(2+\frac{1}{2}).
\end{equation}
We then calculate ${\frac{\langle \triangle H \rangle}{\langle H
\rangle}}$ to obtain
\begin{equation}
\frac{\langle \triangle H \rangle}{\langle H \rangle} = 2.3936m-1.
\end{equation}

The ${\frac{\delta T}{T} \approx 10^{-5}}$ should be given by
${\frac{\langle \triangle H \rangle}{\langle H \rangle}}$. From this
we get ${m}$ by using the density fluctuation values from
observations.

\subsection{Inflation with cosmological constant as dark energy}
The current observations show that our universe is under a late
acceleration phase attributed to dark energy which has an equation
of state ${p=-\rho}$, the only locally lorentz covariant form consistent with the observed
late acceleration phase of the universe. This requires us to include dark energy into
cosmological models. We propose here a model where along with the
false vacuum energy state towards which the universe cools as ${T
\rightarrow 0}$, we have an additional potential in the universe
because of the vacuum fluctuations in space-time. We take the
expectation value of the stress-energy tensor, ${\langle T_{\mu \nu} \rangle_{vac} \equiv \Lambda g_{\mu \nu}}$ to be the cosmological
constant which results in an addition of ${\frac{\Lambda}{8\pi G}}$ to
energy density.

Therefore Eq.~(\ref{thermoenergydensity}) for energy density becomes

\begin{equation}
\rho(T) = \frac{\pi^2}{30} N_1 T^4 + \rho_o + \frac{\Lambda}{8\pi
G}.
\end{equation}
Therefore Eq.~(\ref{frw2}) now takes the form
\begin{equation}
(\frac{\dot{T}}{T})^2 + \epsilon(T)T^2 = \frac{4\pi^3}{45}G N_1 T^4 +
\frac{8\pi}{3}G \rho_o + \frac{\Lambda}{3}. \label{frw-2oic}
\end{equation}

Calculations similar to that shown for classical inflation we get
two solutions based on the value of ${\epsilon}$. If ${\epsilon >
\epsilon_o}$, where

\begin{equation}
\epsilon_o = \frac{8\pi^2}{45}\sqrt{30} G\sqrt{N_1 (\rho_o +
\frac{\Lambda}{8\pi G})}
\end{equation}
Then the universe will have an expanding phase which halts when
temperature reaches ${T_{min}}$ given by
\begin{equation}
T_{min}^4 = \frac{30}{\pi^2}(\rho_o + \frac{\Lambda}{8\pi G})
[\frac{\epsilon}{\epsilon_o} + \{(\frac{\epsilon}{\epsilon_o})^2 -
1\}^{\frac{1}{2}}]^2.
\end{equation}
After this, the universe starts to contract.

For the case when ${\epsilon < \epsilon_o}$, we again consider it to
mainly be ${\epsilon < 0}$ since ${\epsilon_o \simeq 0}$. In this
case we get

\begin{equation}
T(t) = Const. e^{-\chi t},
\end{equation}
where we now have
\begin{equation}
\chi^2 = \frac{8\pi}{3}G\rho_o + \frac{\Lambda}{3}.
\end{equation}

and

\begin{equation}
a(t) = Const. e^{+\sqrt{\frac{8\pi}{3}G\rho_o + \frac{\Lambda}{3}} \
t}.
\end{equation}
In early universe, the ${\rho_o}$ term dominates resulting in
inflation. Whereas in late universe, it is the cosmological constant
term which has a role to play resulting in slow acceleration phase
of the universe which we are observing.

\subsubsection{Evolution of universe in the new model} In Guth's
original work with classical inflation, the initial false vacuum
state of energy density ${\rho_o}$ drives the inflation of the early
universe. However, as the universe evolves ${\rho_o}$ relaxes and
decreases in value whereas the density of the universe increases
bringing inflation to a halt when the acceleration due to false
vacuum energy density is overcome by the deceleration due to the
density increase of the universe.

However, in our model, initially since the value of the cosmological
constant is very small, its the ${\rho_o}$ term which dominates and
guides the expansion. Again as ${\rho_o}$ decreases and density
increases, inflation comes to a halt. But here, as time evolves
though the false vacuum state decays to zero, the cosmological
constant term still remains though very small. When the density of
the universe falls below certain critical value due to the expansion
of the universe, the ${\Lambda}$ term again begins to dominate and
drives the universe to an accelerating phase. But the small value of
${\Lambda}$ implies that this acceleration needs to be very small
compared to that during the initial inflationary phase of the early
universe.

The current observations about our universe suggest exactly similar
behavior about our universe and this strongly prompts us to consider
a cosmological model with the cosmological constant included.

\section{Results and Findings}

The constants and coefficients in our inflationary model using
doublet scalar fields are calibrated using the observational values
about
\begin{itemize}
    \item The composition of normal and dark matter in the universe.
    \item The mass ratio of neutralino and proton.
    \item The WMAP results for density fluctuations in early
    universe.
\end{itemize}

We obtain the following constraints on our parameters in our model:
\begin{enumerate}
    \item ${\lambda_2 \leq 54.83 \times 10^{-15}}$ for the scalar field giving rise to Normal
    Matter in ${\lambda \phi^4}$ model.
    \item ${\lambda_1 \leq 16.586 \times 10^{-17}}$ for the scalar
    field giving rise to Dark Matter in ${\lambda \phi^4}$ model.
    \item ${m_2 \leq 2.5583 \times 10^{-7} M_P}$ for the scalar field giving rise to Normal Matter in ${m^2 \phi^2}$ model.
    \item ${m_1 \leq 1.407065 \times 10^{-8} M_P}$ for the scalar field giving rise to Dark Matter in ${m^2 \phi^2}$ model.
\end{enumerate}

Further, we see that the remnant scalar field doublet at the end of
inflationary epoch provides a starting platform for the appearance
of Higgs and Higgsino fields in the post-inflationary era. Also the
Higgs field potential,
\begin{equation}
V = V_o - \frac{1}{2}\mu^2 \phi^2 + \frac{1}{4}\lambda \phi^4
\end{equation}
can be generated by a linear combination of the two potentials
discussed with ${m^2 \rightarrow -\mu^2}$. The mass of particles
will be proportional to the mass of the Higgs boson and the term
${\sqrt{\frac{\mu^2}{\lambda}}}$.

Thus we have a model which consistently provides some of the initial
elements required in the post inflationary epoch.

\section{Conclusions} In this paper, we have undertaken a detailed
study of standard and modern inflationary models. Analysis of recent
cosmological observations and the WMAP results have created a need
to revise and develop improved models for inflation which gives an
integrated answer to the questions of dark-matter, dark energy and
the cosmological constant problem.

We propose here a model built on simple assumptions which allow us
to include supersymmetry into inflationary models and hence provide
an explanation for the observed dark matter to normal matter ratio.
We go by the hypothesis that supersymmetric partners of normal
matter particles are the best candidates for dark matter and proceed
to show how inflation asymmetrically produces normal and
supersymmetric particles during the inflationary phase.

We fix the limits on the coefficients in our model by calculating
the density fluctuation predicted by our model and matching it with
the observed value of ${10^{-5}}$. Further, we include the
cosmological constant in classical inflation model as a correction
term to the false vacuum energy state which drives inflation.
Because the value of cosmological constant is small, this correction
term is negligible in early universe when both the false vacuum
state as well as the energy density of the universe were large. But
as the density of the universe falls due to the expansion of the
universe, we have the cosmological constant term dominating the
evolution of our universe and hence resulting in a slow late
acceleration of the universe.

Though we develop our model on the assumptions of a doublet primal
scalar field, supersymmetric particles as the best dark matter
candidate and cosmological constant as dark energy, all these
assumptions are supported by similar symmetry existing in related
branches of high energy physics and the recent observational
evidences about the evolution of universe.

However, the nature of connection between the scalar fields of
inflation and the Higgs field is an open question which might need
LHC Higgs mass and astroparticle physics data and models. More
precise data is expected from the PLANCK mission as well as the LHC
project which will help in further narrowing the range of reasonable
cosmological models.

We thank HBCSE, TIFR for giving us the opportunity to work on NIUS
project.

{\centerline {{amruth.br@gmail.com }${ \  \  \  \  \ }$
{aypn1@yahoo.co.in } }}

\end{document}